# Dynamic transverse susceptibility in *Au-Fe-Au* "nano-onions"

H. Srikanth[1], E. E. Carpenter and C. J. O'Connor

*Advanced Materials Research Institute, University of New Orleans, New Orleans, LA 70148*

A precision radio-frequency (RF) transverse modulation technique based on a resonant tunnel diode oscillator (TDO) has been used to study the dynamic susceptibility in novel ring-shaped single domain Fe nanoparticles synthesized using reverse-micelle chemistry. Hysteresis loops measured using a SQUID magnetometer indicate a large coercivity ($H_c$ = 400Oe) below the blocking temperature ($T_B$ ~52K). The dynamic susceptibility shows remarkable consistency with the M-H data and exhibits peaks at characteristic anisotropy fields. Overall, these RF experiments are well suited to probe the magnetic anisotropy in nanophase materials.

Magnetic nanoparticles are materials of great current interest both in terms of remarkable fundamental properties exhibited by these systems as well as their technological potential in the area of high density magnetic storage media [1]. Some of the exciting developments arising from research on single-domain magnetic nanoparticles include observation of macroscopic quantum tunneling (MQT) in antiferromagnetic (AFM) systems [2] and achievement of promising large coercivities in ferromagnetic (FM), AFM and ferrite nanoparticles [3].

Strong exchange interaction in single-domain nanoparticles align the magnetic moments parallel to each other leading to the description of an individual particle itself as a superspin with a giant magnetic moment. In such a system, coherent magnetization rotation is expected to be the dominant mechanism by which these moments align in response to an applied magnetic field. The magnetic axis of a single-domain particle is determined by the magnetic anisotropy energy ($U=KV$) of the particle where $K$ is the total anisotropy energy per unit volume and $V$ is the volume of the particle. At high temperatures, the magnetic anisotropy energy barriers of single-domain particles are easily overcome by thermal energy resulting in superparamagnetism [4]. Below a characteristic blocking temperature ($T_B$), the magnetic axes of the particles are randomly oriented and frozen and under a large external field, all the magnetic axes are aligned giving rise to the saturation magnetization ($M_s$).

Since magnetic anisotropy is the prominent aspect governing the magnetization switching process in nanoparticles, it is important to be able to directly probe its manifestation in the field dependence. Several methods like hysteresis (M-H) loops, Mossbauer spectroscopy, ferromagnetic resonance (FMR), Brillouin light scattering (BLS) have been used to characterize the magnetic anisotropy. While all these experimental methods can yield information, transverse modulation techniques are particularly attractive for direct measurements of the effective anisotropy fields ($H_k$) and the overall distribution of anisotropy fields in nanoparticle aggregates. In a recent paper, Cowburn et al. [5] have reported measurements of the magnetic energy surface in an array of square nanomagnets fabricated by electron beam lithography using modulated field magneto-optical anisometry (MFMA). In particular, the authors have rightly pointed out the significance of transverse modulation techniques in determining the magnetic anisotropy field ($H_k$) from peaks in the transverse susceptibility ($\chi_t$).

We have recently developed a very sensitive way to probe the dynamic susceptibility in the radio-frequency range. Our experiments are based on a novel tunnel-diode oscillator (TDO) technique that combines the advantages of the precision afforded by resonance frequency measurements along with the transverse modulation principle [6]. In this Letter, we present results on nanophase materials containing single-domain ring-shaped *Fe* nanostructures synthesized using sequential reverse micelle processes. Our data provides direct evidence of peaks in the dynamic transverse susceptibility ($\chi_t$) at characteristic fields and also demonstrates a unique way to study the approach to saturation as the static magnetic field is increased.

Nearly uniform spherical nanoparticles having a *Au-Fe-Au* structure with a gold core, ring-shaped annular iron nanostructure surrounded by a gold shell were synthesized. This structure is illustrated in the schematic shown in Fig. 1 and was confirmed from TEM images. Due to the onion-like structure, we refer to these samples as "nano-onions". Note that nanometer-sized Fe particles are pyrophoric and the passivating outer *Au* shell literally protects the material from going up in smoke when exposed to atmosphere.

The nano-onions were synthesized in reverse micelles of cetyltrimethylammonium bromide (CTAB), using 1-butanol as the co-surfactant, and octane as the oil phase [7]. To this mixture, a water solution containing metal ions at a concentration of 0.1M were added. The size of the reverse micelle is determined by the molar ratio of water to surfactant. In the synthesis of the gold core, two micelle solutions were prepared, the first one contained 0.1M Gold (Au3+) (aqueous) and a second one containing sodium borohydride. The two solutions were mixed forming the characteristic red solution. This was stirred for 2 hours for complete reduction of the ions to the metallic state. The micelle was then expanded to allow for a 1nm coating of iron using an aqueous sodium borohydride micelle solution and a fourth micelle solution containing Fe2+ (aqueous). This solution turned black upon adding the Fe2+, and was allowed to react for 2 hours. The micelle was further expanded to form a two nm coating of gold by using a third sodium borohydride solution, and a second aqueous gold solution. The micelle solution was disrupted by adding excess amount of CHCl3:methanol and the nanoparticles were removed using a permanent magnet. The surfactant was removed with successive washings with CHCl3:methanol. The resulting powder containing the aggregate of *Au-Fe-Au* nano-onions, was black in color. Note that the reduction of Au is carried out in the presence of UV light and the entire reaction is performed under Ar gas to minimize oxidation.

Structural characterization of the nano-onions was done using X-ray diffraction and electron microscopy. The particle-size distribution was observed to be within 8%. A SQUID magnetometer (Quantum Design) was used to characterize the magnetic properties. Details of the structural and magnetic characterization will be presented in a separate publication [8].

Examination of the zero-field cooled (ZFC) and field-cooled (FC) susceptibility revealed the characteristic signature of spin-freezing below $T_B = 52K$ and superparamagnetism for $T > T_B$. Field dependent magnetization (M-H) scans displayed no hysteresis in the superparamagnetic state and large co-ercivities at low temperatures below $T_B$. The M-H loop at T = 10K is shown in Fig. 2. A coercive field ($H_c$) of 400 *Oe* and saturation magnetization ($M_s$) around 11.5 emu/mol are obtained.

The RF dynamic susceptibility measurements were done using a *LC*-tank resonator driven by a tunnel diode forward biased in its negative resistance region. This idea is the basis of the TDO measurement system whose circuit design and operation have been described elsewhere in detail [6]. The circuit is self-resonant with a typical resonance frequency around 5 MHz.

The *Au-Fe-Au* nano-onions in powder form are placed in gelcaps that snugly fit into the core of the inductive copper coil (*L*). This is inserted into the sample space of a commercial Physical Property Measurement System (PPMS) from Quantum Design using a customized radio-frequency (*RF*) co-axial probe. The temperature (*T*) and static magnetic field (*H*) are varied using the PPMS. The oscillating RF field ($H_{rf}$) produced by the *RF* current flowing in the coil windings, is oriented perpendicular to the static field *H* and this arrangement sets up the transverse modulation geometry. In the experiment, the measured quantity is the shift in resonant frequency as the static field is varied. The frequency shift ($\Delta\omega$) arises from a change in coil inductance ($\Delta L$) and that in turn gives an accurate measure of the variation in the real part of the transverse susceptibility, $\Delta\chi_t$.

The measured change in transverse susceptibility, $[\chi_t(0)-\chi_t(H)]/\chi_t(0)$, as H is varied upto 15kOe is shown in Fig. 3. This is at a temperature of 10K which is well below the blocking temperature. Peaks in the data, located roughly symmetric about zero can be seen at low fields and the overall variation resembles a bell-shaped curve. Hysteresis can also be clearly made out in the vicinity of the low field peaks. Note that the high density of data points and smooth nature of the curve showcases the precision achieved by our RF method that is not obtained in other experiments. This can directly be ascribed to the nature of this resonance technique and the high sensitivity, ie. the ability to detect changes of a few Hz in 5 MHz.

To map out the low-field hysteresis and study the peaks in finer detail, we started at zero field and cycled H up and down between 1500 and –1500 Oe. The dynamic susceptibility change is plotted in Fig. 4.

The peak structure and hysteresis systematics are now clearly visible. As *H* is varied over the entire loop cycle, two observations can be made about the peak structure – (a) symmetric location around ±0.5kOe and (b) asymmetric peak heights as the field is swept in one direction from -1.5kOe → 1.5kOe or vice versa (follow arrows in Fig. 4). Both these features are remarkably consistent with the (M-H) data from SQUID measurements, shown in Fig. 2. As to the asymmetry in peak heights, the argument can be made as follows. Since the dynamic susceptibility is equivalent to the derivative of magnetization, this is related to the slope of the M-H loop shown in Fig. 2. When the field is cycled from positive to negative, as a consequence of remanence, the segment of the M-H curve in the first quadrant (H > 0) is shorter with a smaller slope than the section in the second and third quadrants (H < 0). Likewise, when the field is changed from negative to positive, the third quadrant segment (H < 0) is shorter than the one in the fourth and first quadrants (H > 0) as zero field is crossed. This asymmetry is directly reflected in the asymmetric peak heights seen in the RF data.

We identify the location of the peaks with the effective anisotropy field ($H_k = 2K/M_s$). The theoretical basis for this can be outlined as follows.

The size of the nano-onions are well below the criterion for single-domain limit and to a good approximation can be considered as ideal Stoner-Wohlfarth (SW) particles [9]. In SW theory for spherical single-domain nanoparticles of uniaxial anisotropy, the magnetization reversal is dominated by coherent rotation. The magnetization energy is given by,

$$E(\boldsymbol{q}) = K\sin^2\boldsymbol{q} \quad (1)$$

where θ is the angle between the magnetization vector and the magnetic axis and K is the energy with which it is bound to this axis. The rigidity of the magnetization against rotations from this axis is,

$$d^2E/d\theta^2 = 2K \qquad (2)$$

This rigidity is considered as being caused by an effective field called the anisotropy field, $H_K$ $(=2K/M_s)$.

In our present transverse modulation geometry, we should consider two torques acting on particle moments due to the static field ($H$) and the orthogonal RF field ($H_{rf}$). Taking into account the Zeeman contributions due to the two fields, the total energy density can be written as [10],

$$U(\theta) = E(\theta) - M_sH\cos(\theta-\phi) - M_sH_{rf}\sin(\theta-\phi) \qquad (3)$$

where φ is the angle between the magnetization vector and H. Following SW formulation, the equilibrium spin configuration is obtained by minimizing eqn. (3) setting $\partial U/\partial \theta = 0$. Also, since H is large with respect to $H_{rf}$, the small angle approximation can be made as (θ–φ) ~ 0. This is equivalent to stating that the spins are closely aligned with the static magnetic field. This leads to the equation,

$$dE/d\theta = M_sH_{rf} - M_sH(\theta-\phi) \qquad (4)$$

Differentiating equation (4) with respect to θ and inverting, we get

$$d\theta/dH_{rf} = [(d^2E/d\theta^2)/M_s + H]^{-1} \qquad (5)$$

The left hand side can be identified as the change in transverse susceptibility ($\chi_t$) which is exactly the quantity being measured in the experiment. From expression (2), the first term on the right hand side of (5) is easily identified as the anisotropy field $H_K$. The singularity evident from equation (5) accounts for the peak structure seen in the experimental data of Fig. 4 as the swept static field crosses the effective anisotropy field.

The $H_K$ values of around *450 to 500Oe* obtained with the TDO experiments are only slightly larger than the co-ercivity of ~ *400Oe* inferred from magnetization measurements. This is to be expected in the case of single-domain particles with uniaxial anisotropy, as the co-ercivity ($H_c$) in the SW framework is given by [11],

$$H_c = H_K[1-A/D^{3/2}] \qquad (6)$$

where D is the diameter of the nanoparticle and A is a factor dependent on temperature, *K*, measurement time, attempt frequency.

The distribution in anisotropy fields can be studied by fitting the symmetric bell-shaped curve of Fig. 3 using various distribution functions [12]. Our attempts to fit using log-normal, Gaussian and Lorentzian distribution functions indicated that Lorentzian or Cauchy distributions generated the best fit nearly indistinguishable from the experimental data. It is interesting to note that cumulative distributions can be numerically computed from the dynamic susceptibility data, which will generate curves resembling (M-H) variation under the influence of two orthogonal fields, H and $H_{rf}$. Detailed studies of the distributions obtained from RF transverse modulation experiments on single-domain nanoparticles as a function of particle sizes and comparison with larger-grain multi-domain systems, are currently under way and will be presented in a forthcoming publication [13].

Finally, the dynamic susceptibilty data at different temperatures are shown in Fig. 5. The double peak structure disappears well above the blocking temperature of 52K deduced from DC susceptibility measurements. A trace of the structure is present at temperatures above $T_B$. Since the TDO experimental frequency is around 5MHz, this could well signify that the blocking temperature is frequency dependent and that would then be a strong indication for non-interacting particles [14]. Further experiments are needed to study the role of interactions and magnetic relaxtion effects at radio frequencies.

This research is supported by DARPA through grant No. MDA 972-97-1-0003. The authors would like to thank Jason Wiggins for assistance with RF instrumentation. The contributions by AMRI members Weilie Zhou, Joan Wiemann and Claudio Sangregorio towards structural and magnetic characterization of the samples, are duly acknowledged. The authors also thank Jinke Tang for his critical reading of the manuscript and useful comments.

[1] Corresponding author; e-mail: sharihar@uno.edu

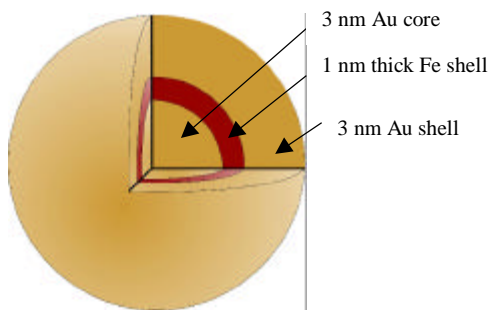

Fig. 1 Schematic of the *Au-Fe-Au* "nano-onion".

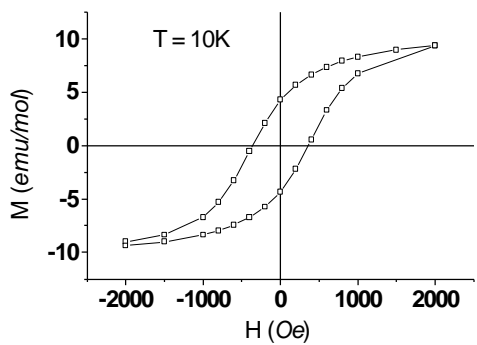

Fig. 2 Hysteresis loop measured with a SQUID magnetometer

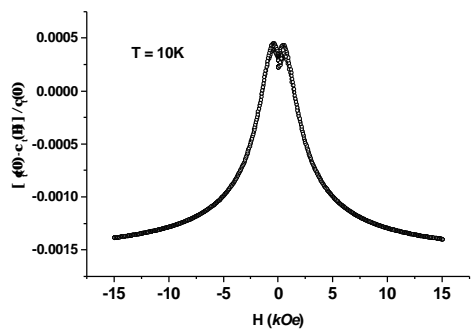

Fig. 3 Dynamic transverse susceptibility of the Au-Fe-Au nano-onions measured using the RF TDO method

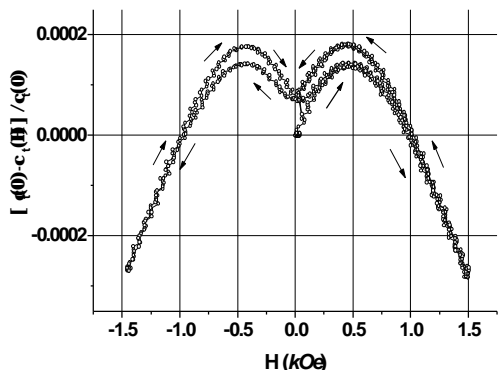

Fig. 4 Low-field hysteresis and peak structure in the dynamic susceptibility. The arrows represent the direction as the field is ramped through a full cycle: 0 → 1.5kOe → -1.5kOe → 1.5kOe → 0

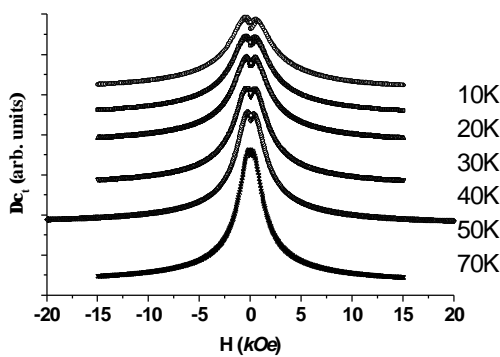

Fig. 5 Dynamic transverse susceptibility at various temperatures. Data sets are relatively shifted for clarity.